# Deep-tissue large field of view imaging by Fourier conjugate adaptive optics

LYUBOV V. AMITONOVA

*Biomedical Photonic Imaging Group, University of Twente, P.O. Box 217, 7500 AE Enschede, The Netherlands*

*LaserLaB, Department of Physics and Astronomy, Vrije Universiteit Amsterdam, De Boelelaan 1081, 1081 HV Amsterdam, The Netherlands*
*Corresponding author: l.amitonova@vu.nl*

**Light microscopy enables multifunctional imaging of biological specimens at unprecedented depths and resolutions. However, the performance of all optical methods degrades with the imaging depth due to sample-induced aberrations. Methods of adaptive optics (AO) are aimed at pre-compensation of these distortions but state-of-the-art adaptive optics still provides a limited field of view and imaging depth. Here I propose a new approach to overcome these limitations: Fourier image plane conjugate AO. Two possible experimental designs of the new approach are investigated and an accurate comparison between proposed and previously used methods of AO is presented. We see that Fourier conjugate AO provides a significantly larger field of view, which can only be limited by the angular optical memory effect, as well as it is simpler in practical realization for large imaging depth and allows the optimal use of resolution of spatial light modulator.**

Light microscopy has been a key tool for biological and medical research for centuries [1]. Optical approaches can provide high spatial resolution together with extended functionality produced by the number of different contrast mechanisms [2,3]. However, light scattering in biological tissues restricts *in vivo* application of these techniques to the near-surface region [4]. Nowadays, the most popular strategy to overcome this limit is to use only ballistic photons. Optical coherence tomography, confocal microscopy, and nonlinear optical microscopy exploit or partially exploit this scenario. Unfortunately, the number of ballistic photons decays exponentially with depth (following the Beer-Lambert law) leading to an enormous drop in the signal after several mean free paths of a photon [4]. As a result, the imaging depth is still limited to approximately 1 mm even in state-of-the-art works [5,6]. Currently, only invasive fiber probes can be used for high-resolution deep-tissue optical imaging [7,8].

The next breakthrough on the way to noninvasive deep-tissue *in vivo* optical microscopy is the implementation of an advanced adaptive optics (AO) technique, which is based on standard AO microscopy and the wavefront shaping [9]. Adaptive optics is typically used to compensate the known system aberrations [10,11], whereas the wavefront shaping technique [12–14] allows the creation of the desired intensity profile behind the highly scattering sample [15]. The concept of wavefront engineering combined with advanced optical microscopy methods is the only way to look deep inside highly scattering biotissues [16]. The most common AO approach involves placing a deformable mirror (DM) or spatial light modulator (SLM) in the pupil plane of the imaging device, which is a standard telescope in 4f configuration [17]. Wavefront correction is designed to compensate the aberrations only at a particular spot with the hope that the maintained degree of correlation between initial and scanned beams – optical memory effect [18,19] – will be large enough to visualize neighbouring objects. However, in this scenario the so-called shift memory effect is utilized, which is very small even for a single scattering layer. This problem can be partially solved by multi-pupil adaptive optics, but this approach requires a complex setup for realization [20].

More promising is the use of an angular optical memory effect, which requires a conjugate AO microscope configuration: placing the DM or SLM conjugate to the main source of aberrations. The conjugate AO is found to provide a significant field of view advantage [21,22]. The main principle of a typical optical setup for conjugate AO is presented in Fig. 1(a) where the SLM is placed in the plane where the image of scattering layer with lens system $\{L_2\,L_3\}$ is formed. The single scattering layer is represented by a solid black curve. The optimal wavefront for zero input angle is projected on the SLM as shown in Fig. 1(a) by the solid blue curve. The angular optical memory effect allows us to use part of the previously optimized wavefront during the scanning procedure. It is assumed that aberrations in a parallel beam are fully compensated. We also assume that pre-compensation was performed for a single scan angle, which is typical for a strongly scattering layer [23]. As a result, for non-zero input angles, the pre-compensation still partially works – the white part of the black curved line after lens $L_3$.

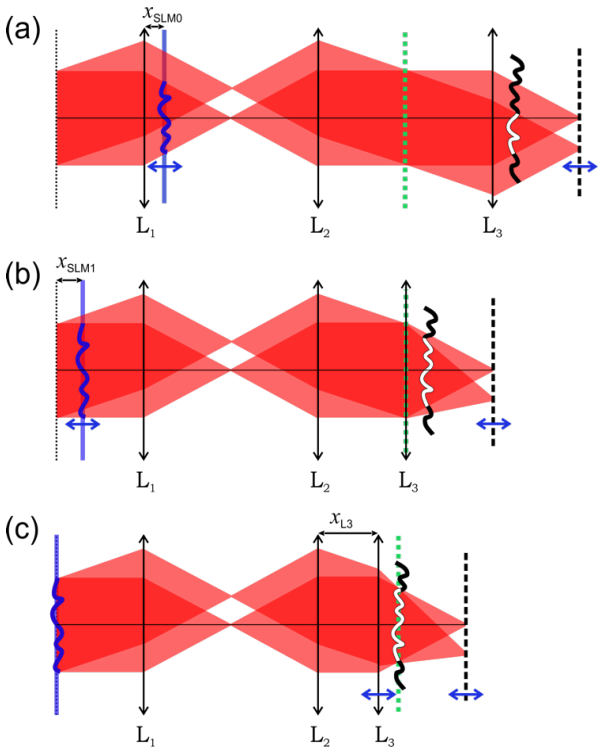

Fig. 1. Illustration of the main ideas of different conjugate AO methods: (a) standard conjugate AO and (b-c) two designs of proposed Fourier conjugate AO. Lenses $L_1$, $L_2$, and $L_3$ have the same focal distances. Dotted green lines represent the pupil planes. Single plane phase distortion is shown by the black curved line after lens $L_3$. The image plane and the SLM position is shown by the blue line. The distortion is fully pre-compensated by the optimal wavefront for zero input angle. For non-zero input angles, the pre-compensation works only partially – the white part of the black curved line after lens $L_3$. We already see that for the same parameters of the optical system and the same input angles, pre-compensation works better for large angles in the case of Fourier conjugate AO.

The efficiency of conjugate AO approach was demonstrated in recent experiments [24–28]. Conjugate AO allowed high-resolution *in vivo* imaging of a mouse brain through the intact skull [29]. However, the field of view is still quite limited and despite all potential advantages for deep-tissue microscopy, the imaging depth typically doesn't exceed 1 mm. On top of this, to place the DM or SLM in the conjugate plane in the focusing beam is an experimental challenge, especially for high imaging depths [21].

Here I propose and investigate a more powerful method of adaptive optics: Fourier image plane conjugate AO. This new method has several advantages: (1) it provides high FOV within the memory effect range even if pre-compensation was done only for a single focal point; (2) it is more appropriate for large imaging depth and (3) it doesn't require putting the SLM in the focusing beam, which allows the optimal use of SLM resolution and excludes its possible thermal damage.

The illustration of the main principle of Fourier conjugate AO is presented in Fig. 1(b, c). The spatial light modulator is placed in the plane where the image of the scattering layer with the lens system {$L_1 L_2 L_3$} is formed, the so-called Fourier image plane. This new approach allows the realization of different microscope designs.

In the first design, lens $L_3$ is fixed and placed into the focal plane of lens $L_2$ (see Fig. 1(b)). The position of the SLM depends on the position of the scattering layer. The optimal wavefront for zero input angle is projected on the SLM as shown in Fig. 1(b) by solid blue line and aberrations in a parallel beam (guided star beam) are fully compensated. While scanning, the aberration is partially compensated (white part of the distorted phase), depending on the scan angle. We easily see that the compensated area in Fourier conjugate AO is much higher than in standard conjugate AO for the same parameters of optics and scan angles (compare white solid curves in Fig. 2(a) and Fig. 2(b)).

In the second design of Fourier conjugate AO, lens $L_3$ is placed in such a way that a 'pupil' plane at the position of the scattering layer is formed. The SLM is placed into the back focal plane of lens $L_1$ and its position is fixed (see Fig. 1(c)). For proper conjugation the position of lens $L_3$ should be adjusted. We see that this design allows to maintain the full compensation of the aberrations independently of the scan angle (beam is always within a pre-compensation area – white curve in Fig. 1(c)). Resultantly, in this microscope configuration FOV is limited only by angular memory effect range, which is infinite for a single scattering layer.

All three approaches were analytically characterized by using the thin lens equation in the paraxial ray approximation in terms of the best possible FOV, imaging depths, and optimal microscope parameters. Numerical simulations of beam propagations were then made by using the beam propagation method to confirm theoretical predictions.

Firstly, we analyze the difference in field of view for standard conjugate AO and Fourier conjugate AO in the case of a single scattering layer and single focus pre-compensation. Here potential FOV is estimated as a distance at which focusing is done with half of the wavefront along this axis fully compensated. We can see that for standard conjugate adaptive optics (CAO) field of view is given by $FOV_{CAO} \propto D \cdot f_2 \cdot l/(f_1 \cdot f_3)$; and for the first configuration of the Fourier conjugate adaptive optics (FCAO) by $FOV_{FCAO} \propto D \cdot f_2 \cdot l/(f_1 \cdot (f_3 - l))$, where $f_1$, $f_2$ and $f_3$ are focal lengths of $L_1$, $L_2$ and $L_3$ lenses respectively, $l$ is distance between the scattering layer and the focal plane (imaging depth beyond the scattering layer, see Fig. 1(a-b)) and $D$ is the input beam diameter. The ratio between $FOV_{FCAO}$ and $FOV_{CAO}$ as a function of normalized imaging depth beyond the scattering layer, $l/f_3$, is plotted in Fig. 2(a). We see the significant enhancement of field of view in Fourier conjugate AO, especially for large imaging depths. Moreover, the second configuration of Fourier conjugate AO, which is presented in Fig. 1(c), provides imaging with FOV limited only by the angular optical memory effect.

Secondly, we calculate the optimal positions of the SLM for the setups presented in Fig. 1(a-b) and optimal position of the $L_3$ lens for the setup presented in Fig. 1(c). As a result, the position of the SLM for the standard conjugate AO is given by $x_{SLM0} = f_1 - l \cdot (f_2/f_3)^2$; the position of the SLM for the Fourier conjugate AO in the first configuration is given by $x_{SLM1} = f_3 \cdot (f_1/f_2)^2 (f_3/l - 1)$ and the position of lens $L_3$ for the Fourier conjugate AO in the second configuration is given by $x_{L3} = f_2 - f_3(f_3/l - 1)$. By using these data, we can characterize the flexibility of the microscope configurations and find the imaging depth limits for all microscope setups. As a result, $l/f_3 < 1/p$ for standard conjugate AO, $l/f_3 > 1/(p + 1)$ for the first configuration (see Fig. 1(b)) and $l/f_3 > 1/(f_2/f_3 + 1)$ for the second configuration (see Fig. 1(c)) of Fourier conjugate AO, where $p = f_2^2/(f_1 f_3)$ is a normalized parameter of an optical system.

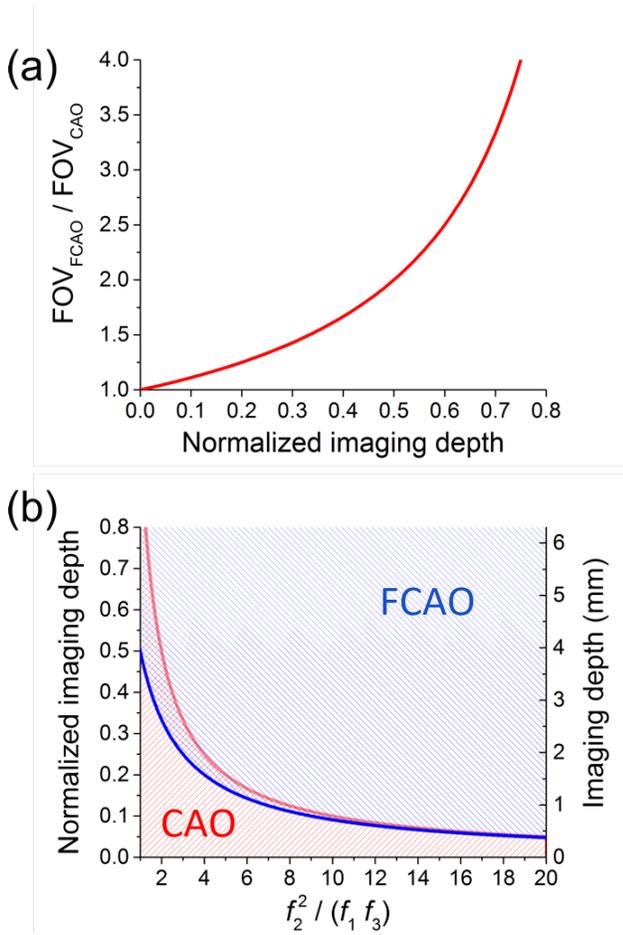

Fig. 2. (a) FOV advantage of Fourier conjugate AO. The ratio between FOV in the second configuration (Fourier conjugate AO (FCAO), see Fig. 1(b)) and FOV in the first configuration (standard conjugate AO (CAO), see Fig. 1(a)) as a function of normalized imaging depth beyond the scattering layer $l/f_3$. (b) Limits of possible normalized imaging depth beyond the scattering layer, $l/f_3$, as a function of $p$ parameter of the optical system, $p = f_2^2/(f_1 f_3)$, for standard conjugate AO (red line) and Fourier conjugate AO (blue line). Possible imaging depths are situated below the red curve for standard conjugate AO (red dashed area) and above the blue curve for Fourier conjugate AO (blue dashed area). The right axis shows the imaging depth in mm for typical microscope objective with $f_3 = 9$ mm. Limit for the second configuration of Fourier conjugate AO is equal to limit of the first configuration for the case of $f_1 = f_2$.

The results are presented in Fig. 2(b) where the limitations of normalized imaging depths beyond the scattering layer, $l/f_3$, are plotted as functions of the $p$ parameter of a system. The red line corresponds to the limit for standard conjugate AO and the blue line represents a limit for the first configuration of Fourier conjugate AO. Theoretically possible imaging depths lie below the red curve for standard conjugate AO (red dashed area in Fig. 2(b)) and above the blue curve for Fourier conjugate AO (blue dashed area in Fig. 2(b)). The limit for the second configuration of Fourier conjugate AO is equal to the limit of the first configuration for the case of $f_1 = f_2$.

We see that the two methods of conjugate AO suffer from different limitations. Standard conjugate AO always works for small imaging depths but for increasing imaging depth we need to move DM or SLM closer to lens $L_1$, placing thereby a theoretical limit on the parameters of the optical system and reachable imaging depths.

In contrast, Fourier conjugate AO always works for large imaging depth, with the limit arising as imaging depth is lowered. For decreasing imaging depth, we need to move the SLM closer to lens $L_1$ in the first Fourier conjugate AO configuration (see Fig. 1(b)) and move lens $L_3$ closer to lens $L_2$ in the second Fourier conjugate AO configuration (see Fig. 1(c)).

Despite the fact that both approaches have parameters, which allow to use the full depth range, we see that Fourier conjugate AO is more appropriate for deep tissue imaging. Microscopy setups based on conjugate AO require a decrease in focal distance of lens $L_1$ to increase imaging depth. For example, standard conjugate AO microscopes optimized for typical objectives with large working distances, such as UMPLFLN Olympus (20x, NA = 0.5; WD = 3.5 mm) and LMPLFLN Olympus (20x, NA = 0.4; WD = 8 mm) provide a maximal imaging depth of only 1.8 mm and 2 mm respectively. In contrast, Fourier conjugate AO approaches (second and third configurations, Fig. 1(b-c)) provide imaging depth limited only by working distance of the objective. In calculations we take the minimal focal length of lens $L_1$ as 50 mm; SLM size of 10 mm and input beam diameter of 10 mm. As a result, if we aim at deep-tissue imaging it is preferable to use the new method of Fourier conjugate AO.

The final advantage of Fourier conjugate AO is related to practical reasons of convenience and stability of an experimental setup, as well as optimal usage of SLM pixels. In standard conjugate AO approaches, DM or SLM is placed into the focusing beam [21]. It means that for different imaging depths a different number of pixels afor phase compensation is available. It will greatly influence the quality of wavefront pre-compensations for different depth. For Fourier conjugate AO, the beam on the SLM is always parallel and a number of pixels used doesn't depend on imaging depth. Moreover, the second configuration of Fourier conjugate AO setup fixes the SLM position for different imaging depths. This makes Fourier conjugate AO more suitable for 3D visualization.

Numerical simulations of beam propagations were made to confirm theoretical predictions. We simulate focusing through a single scattering layer with the help of standard conjugate AO and two different approaches of Fourier conjugate AO using a beam propagation method [30]. The following parameters were used to simulate all methods (the intersection of blue and red dashed line areas in Fig. 2(b)): focal lengths $f_1 = f_2 = f_3 = 100$ mm, which lead to $p = 1$, Gaussian beam with input beam diameter D = 1 mm, position of scattering layer (imaging depth) $l = 80$ mm, deep-tissue imaging, $l/f_3 = 0.8$. As a single scattering layer, an artificially generated random phase mask presented in Fig. 3(e) were used. For SLM patterns we use the same phase mask properly scaled. On the virtual SLM, only the central part of the phase is projected as if someone characterized it only for the beam propagated at zero angle.

The results of simulations are presented in Fig. 3(a) for a standard conjugate AO approach (experimental setup depicted in Fig. 1(a)) and in Fig.3(b-c) for Fourier conjugate AO approaches (experimental setups depicted in Fig. 1(a-b), respectively). We simulate output beam profiles for different input angles and present sum over focal points distributed with the step 0.2 mm. In Fig. 3(d) we see normalized peak intensity as a function of scanning radius for standard conjugate AO (red line), the first design (blue line) and the second design (green line) of Fourier conjugate AO.

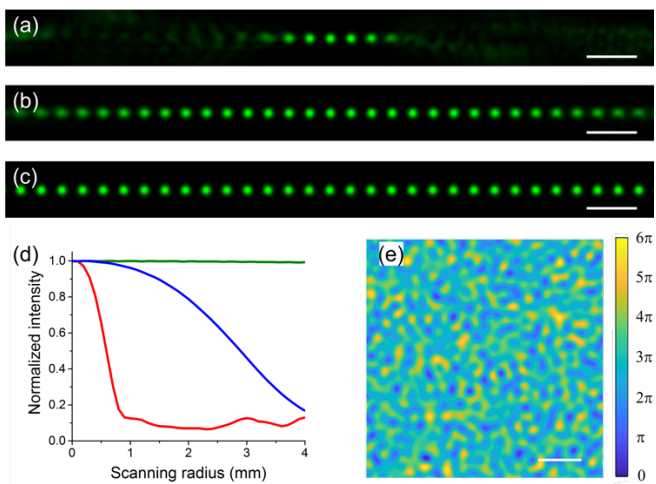

Fig. 3. (a-d) Results of simulated beam profiles on the output for different input angles using a beam propagation method. (a-c) Sums over simulated focal points distributed with a step of 0.2 mm are presented for (a) the standard conjugate AO approach, (b) the first configuration of Fourier conjugate AO approach, (c) the second configuration of Fourier conjugate AO approach. (d) Normalized peak intensity as a function of scanning radius for standard conjugate AO (red line, experimental setup depicted in Fig. 1(a)), the first configuration of Fourier conjugate AO (blue line, experimental setup depicted in Fig. 1(b)) and the second configuration of Fourier conjugate AO (green line, experimental setup depicted in Fig. 1(c)). (e) A single scattering layer used as a random phase mask. Scale bars are 0.5 mm (a-c) and 1 mm (e).

We see that for standard conjugate AO the quality of the focal spot degrades very fast, as predicted by theoretical analysis. Theoretical predictions state that the FOV = 0.8D for standard conjugate AO and FOV = 4D for the first configuration of Fourier conjugate AO for chosen parameters of a system. Our simulations show that at radius r = 0.4D = 0.4 mm from the center, the peak intensity decreases to ~80% of the maximum as well as the quality of focal spot noticeably degrades (see Fig. 3(a)) for the standard conjugate AO approach. In contrast, the peak intensity and the shape of the focal spot maintained high quality for r = 0.4 mm in the case of Fourier conjugate AO. The peak intensity decreases to ~80% of the maximum, while the quality of the focal spot noticeable degrades (see Fig. 3(b)) for Fourier conjugate AO approach only at a distance 2D = 2 mm from the center, as predicted by the theory above. We also see that the FOV for the second configuration of Fourier conjugate AO is not limited because we consider a single scattering layer with an infinite memory effect range (see Fig. 3(c)).

Here analysis was done for a single scattering layer, but it is straightforward that Fourier conjugate AO will maintain all advantages such as larger FOV and the convenient practical setup adapted for high imaging depth also for bulk scattering samples. In the bulk case, the field of view will be mostly limited by the generalized optical memory effect [31]. The overlapping volume between initial beam ('guide star beam') and scanning beam in both cases of Fourier conjugate AO are significantly bigger than the overlapping volume in the case of standard conjugate AO as can be easily seen by comparing the dark red areas after lens $L_3$ in Fig. 1(a) and in Fig. 1(b-c).

To summarize, I proposed and analysed a very powerful method of adaptive optics: Fourier image plane conjugate adaptive optics. The new method provides a higher field of view, which is limited only by the optical memory effect even if pre-compensation was done for a single focal point, is more appropriate for deep-tissue imaging, is easier and more stable in experimental implementation and allows optimal usage of SLM resolution. Advantages of two different experimental configurations of Fourier conjugate AO were demonstrated. The new method can be used as a powerful tool for noninvasive deep tissue optical microscopy.

**Funding.** Nederlandse Organisatie voor Wetenschappelijk Onderzoek (NWO) (Veni grant);

**Acknowledgment**. I thank Ivo Vellekoop and Gerwin Osnabrugge for support and discussions.